\documentclass[aps,prl,twocolumn,showpacs,groupedaddress]{revtex4}
\usepackage{graphicx}

\begin{document}

\title{Macroscopic spin tunneling and quantum critical behavior of a condensate in
double-well potential}

\author{Han Pu, Weiping Zhang and Pierre
Meystre}
\affiliation{Optical Sciences Center, The University of
Arizona, Tucson, AZ 85721}
\date{\today}

\begin{abstract}
In a previous work \cite{pu}, we have shown that a spinor
condensate confined in a periodic or double-well potential
exhibits ferromagnetic behavior due to the magnetic dipole-dipole
interactions between different wells, and in the absence of
external magnetic field, the ground state has a two-fold
degeneracy. In this work, we demonstrate the possibility of
observing macroscopic quantum spin tunneling between these two
degenerate states and show how the tunneling rate critically
depends on the strength of the transverse field.
\end{abstract}
 \maketitle

Tunneling, a process in which a system penetrates into a
classically forbidden region (e.g., a potential barrier), is an
intrinsically quantum effect with no classical counterpart. Of all
tunneling effects, macroscopic quantum tunneling (MQT), the
tunneling of a macroscopic variable of a macroscopic system
\cite{leggett}, represents a particularly intriguing and
interesting scenario as it touches the boundary between the
classical and the quantum world and may help shed light on the
quantum-classical interface. As such, it is one of the most
striking manifestations of quantum mechanics. The tunneling of
large magnetic spins has recently received much attention, both
theoretically and experimentally \cite{book, qtm94} in view of its
promise as one of the few realistic candidates for an experimental
demonstration of MQT, and also because of its connection to
quantum computing. Despite considerable efforts, though, there are
not many clear and definitive demonstrations of MQT in spin
systems so far --- although there have been some indications that
tunneling might be the underlying reason for some observed results
\cite{qtm94,exp}. The reasons for this difficulty are as follows:
First, most theories apply to single system, while the
conventional magnetic materials used in the experiments contain
many domains, each possessing its own set of parameters such as
magnetic anisotropy and barrier energy; second, due to the
difficulty of cooling the samples down to ultracold temperatures,
thermal processes cannot be completely excluded; and third, the
spins in solid materials are inevitably imbedded in a crystal
matrix and the spin-matrix interaction \cite{matrix} complicates
the physical picture. For these reasons, macroscopic spin
tunneling remains one of the most anticipated, yet elusive,
quantum phenomena.

In this paper, we demonstrate the possibility of observing spin
tunneling in a spinor atomic condensate trapped in a double-well
potential, thereby eliminating most of these difficulties. We
remark at the outset that while the inter-well tunneling of
condensates has been previously considered, all previous studies
focused instead on the tunneling of an {\em external} degree of
freedom, the condensate center-of-mass motion
\cite{milburn,smerzi,jessen,zhang}.

In an earlier paper \cite{pu}, we showed that because of the long
range magnetic dipole-dipole interaction, a spinor atomic
condensate trapped in a one-dimensional periodic lattice potential
or a double-well potential behaves as a ferromagnet. In the
absence of external magnetic fields, the ground state spins of the
``mini-condensates'' confined in individual wells all align
parallel to each other and along the lattice direction (the
$z$-axis), giving rise to a spontaneous magnetization along $z$.
For the sake of simplicity, we restrict the present discussion to
a double-well potential, each well containing $N$ spin-1
condensate atoms. In the tight-binding approximation
\cite{milburn}, they are described by the zero-temperature spin
Hamiltonian \cite{pu}
\begin{equation}
H=\lambda_a' ({\bf S}_1^2+{\bf S}_2^2) -3\lambda S_1^z S_2^z +
\lambda {\bf S}_1 \cdot {\bf S}_2 - h (S_1^x + S_2^x),
\label{hspin}
\end{equation}
where ${\bf S}_i$ is the total spin of the condensate in the
$i^{\rm th}$ well, and $S_{i}^{x,y,z}$ its cartesian components.
The first term in (\ref{hspin}) represents the on-site Hamiltonian
of the spinor condensate. It includes short-range nonlinear
spin-exchange interactions \cite{law}, where the parameter
$\lambda_a'$ is related to the $s$-wave scattering lengths
\cite{ho} and needs to be negative. The second and third terms in
(\ref{hspin}) arise from the site-to-site dipole-dipole
interaction \cite{pu}, where $ \lambda \equiv \gamma_B^2
\mu_0/(4\pi r^3)$ for pure magnetic dipolar interaction, with
$\gamma_B$ being the gyromagnetic ratio, $\mu_0$ the vacuum
permeability and $r$ the distance between the two wells. The value
of $\lambda$ can be greatly enhanced by the light-induced optical
dipolar interaction if one chooses appropriate laser fields to
form the potential well \cite{spinwave,optical}. The last term
describes the effect of an external transverse magnetic field,
taken to be along the $x$-axis without the loss of generality.
Here $h=\gamma_B B$, with $B$ being the strength of the applied
field. In the following, we assume that the nonlinear short-range
atom-atom interaction is strong enough that the first term in
Hamiltonian (\ref{hspin}) dominates over the magnetic dipolar
interaction \cite{note}. As a result, the total spin quantum
number for each mini-condensate is fixed to its maximum value $N$
--- the number of particles in each well. We can therefore neglect
the first term in the Hamiltonian (\ref{hspin}), since it is a
constant of motion and commutes with the remaining terms.

\begin{figure}
\includegraphics*[width=8cm,height=4.4cm]{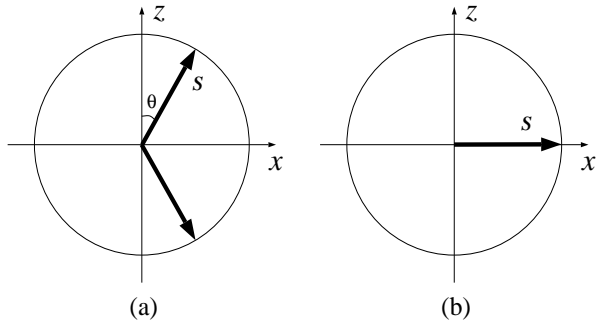}
\caption{Classical ground state spin orientation. (a) For $0 \leq
h < 3N \lambda$, the ground state has a two-fold degeneracy. (b)
For $h > 3N \lambda$, the degeneracy is removed and the spins are
polarized along the transverse field. } \label{fig1}
\end{figure}

Before discussing quantum mechanical spin tunneling, let us first
investigate the classical situation. The Hamiltonian in that limit
is still given by (\ref{hspin}), except that the spins are now
$c$-numbers that can be represented by vectors of fixed length $N$
in spin space. For zero field, $h=0$, it is easy to see that the
classical ground state is two-fold degenerate with ${\bf S}={\bf
S}_1={\bf S}_2$ pointing along either the $z$- or $(-z)$-axis.
Under the influence of a weak transverse field along the
$x$-direction, the two ground states move away from the $z$-axis
and towards the $x$-axis while remaining in the $xz$-plane, as
shown in Fig.~\ref{fig1}. For $0 \leq h <h_c^{(c)}= 3N\lambda$,
the two minima are located at
\[ \theta = \pi/2 \pm \cos^{-1} \left( h/h_c^{(c)}
\right), \]where $\theta$ is the angle between ${\bf S}$ and the
$z$-axis. For $h \geq h_c^{(c)}$, the two minima merge along the
$x$-axis, and the degeneracy is removed. Hence $h_c^{(c)}$ can be
regarded as the classical critical field strength, beyond which
the system is completely polarized by the external field.

Let us now turn to a quantum mechanical description of the system.
Our goal is to investigate whether or not tunneling is present in
the classically degenerate regime (i.e., when $0 \leq h <
h_c^{(c)}$). We will present both a full numerical calculation and
analytical results using the instanton technique.

\begin{figure}
\includegraphics*[width=7cm,height=7.4cm]{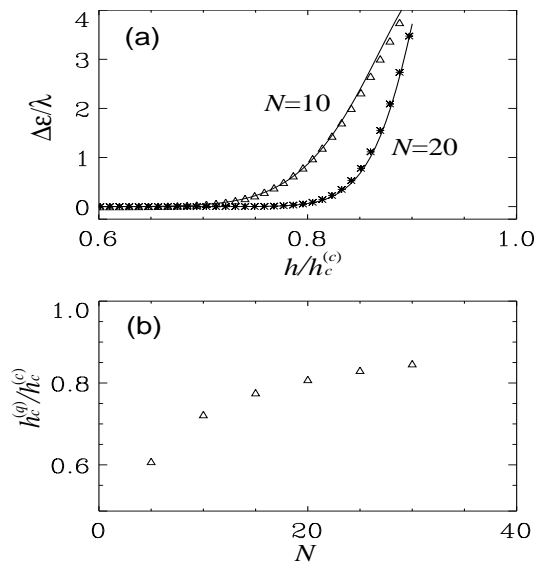}
\caption{(a) Energy splitting $\Delta \varepsilon$ as a function
of the transverse field strength. The symbols represent numerical
results and the solid curves are analytical results obtained from
Eq.~(\ref{instanton}). (b) The quantum critical field strength
$h_c^{(q)}$ as a function of $N$. $h_c^{(q)}$ is defined as the
value of $h$ at which $\Delta \varepsilon=0.1\lambda$. }
\label{fig2}
\end{figure}

A well-known consequence of the tunneling between two degenerate
states is the lifting of their degeneracy: The two new eigenstates
are a symmetric and an antisymmetric superposition of the original
states characterized by an energy difference (or tunneling
splitting) $\Delta \varepsilon$ inversely proportional to the
tunneling rate. The quantity of interest to determine the
occurrence of tunneling is therefore this energy difference
between the two lowest eigenstates of the Hamiltonian. We
determine it by expanding the Hamiltonian (\ref{hspin}) onto the
basis spanned by $S_1^z \otimes S_2^z$, and evaluate numerically
the eigenvalues of the resultant $(2N+1)^2 \times (2N+1)^2$
matrix. Fig.~\ref{fig2}(a) summarizes the result of this analysis.
It shows that $\Delta \varepsilon$ is essentially zero for small
values of $h$, but becomes finite when $h$ exceeds a threshold
value $h_c^{(q)}$ --- the quantum critical field strength. This
means that for $0 \leq h \leq h_c^{(q)}$, quantum mechanics agrees
with classical mechanics in that the system is degenerate.
However, for $h_c^{(q)} < h < h_c^{(c)}$, even though the system
is still degenerate in the classical picture, the presence of
tunneling removes the degeneracy in the quantum treatment.
Fig.~\ref{fig2}(b) displays $h_c^{(q)}$ as a function of $N$, from
which one sees that $h_c^{(q)}$ increases with $N$ and approaches
$h_c^{(c)}$ as $N$ tends to infinity. In other words, as $N$
increases, the system, as expected, behaves more and more
classically.

To gain some analytical insight, we first notice that
Hamiltonian~(\ref{hspin}) can be rewritten as
\[ H= -\frac{3\lambda}{4} [(S^z)^2 - (S^{\prime z})^2] +
\frac{\lambda}{2} {\bf S}^2 - h S^x \] where ${\bf S}={\bf S}_1 +
{\bf S}_2$ and ${\bf S}'={\bf S}_1 - {\bf S}_2$. For
$h<h_c^{(c)}$, ${\bf S}_1$ and ${\bf S}_2$ are tightly bound
together, such that $S^{\prime z} \approx 0$ and ${\bf S}$ is
approximately a constant of motion with quantum number $S=2N$.
Hence, neglecting the constant terms, the effective Hamiltonian of
the system reads
\begin{equation}
H_{\rm eff} = -\frac{3\lambda}{4} (S^z)^2 - hS^x. \label{heff}
\end{equation}
Hamiltonian (\ref{heff}) describes a quantum spin with the
easy-axis anisotropy in a transverse field and it has been
extensively studied in the context of spin tunneling \cite{book}.
Using the instanton technique, which has been proved to be quite
accurate for $S>5$, the tunneling splitting between the two
classically degenerated ground states can be expressed as
\cite{garg}
\begin{equation}
\Delta \varepsilon = p \omega \sqrt{S_c/(2 \pi)}\, e^{-S_c},
\label{instanton}
\end{equation}
where $p$ is a prefactor on the order of unity [by fitting the
numerical results with Eq.~(\ref{instanton}), we find $p \approx
2.75$], $\omega = h_c^{(c)}x$ is the typical instanton frequency
with $x = \sqrt{1-(h/h_c^{(c)})^2}$, and $ S_c = 2N \ln [
(1+x)(1-x)]  -4Nx $ is the classical action. The solid curves in
Fig.~\ref{fig2}(a) represent the values calculated using
Eq.~(\ref{instanton}), which agree well with the numerical
results.

A more intuitive understanding of this threshold behavior in
quantum tunneling can be obtained from a quantum perturbation
viewpoint. Tunneling results from the transverse magnetic field.
If we regard the last term in Hamiltonian~(\ref{hspin}) as a
perturbation to the rest of the Hamiltonian, which possesses two
degenerated ground states $|N,N \rangle$ and $|-N,-N \rangle$,
where we have labelled the states with the eigenvalues of $S_1^z$
and $S_2^z$, the tunneling level splitting is then, in the high
order perturbation theory \cite{qm}, given by the shortest chain
of matrix elements and energy denominators connecting these two
states \cite{garanin}. For the situation at hand, this chain can
be represented as
\begin{widetext}
\begin{eqnarray*}
|N,N\rangle & \Rightarrow  & \frac{1}{\sqrt{2}} (|N-1,N \rangle
+|N,N-1 \rangle) \Rightarrow |N-1,N-1 \rangle \Rightarrow
\frac{1}{\sqrt{2}} (|N-2,N-1 \rangle
+|N-1,N-2 \rangle) \Rightarrow \nonumber \\
&& ... \Rightarrow \frac{1}{\sqrt{2}} (|-N,-N+1 \rangle +|-N+1,-N
\rangle ) \Rightarrow |-N,-N\rangle.
\end{eqnarray*}
\end{widetext}
Since the spin has to travel a distance of $4N$ to reach from one
ground state to the other, the tunneling appears minimally in the
$(4N)^{\rm th}$ order of perturbation theory. The energy splitting
associated with this chain can be calculated as
\begin{eqnarray*}
 \Delta \varepsilon &=& 16 \lambda \left(\frac{h}{2\sqrt{2} \lambda}
 \right)^{4N} \left[ \frac{\Gamma \left(3/2+\sqrt{N^2+1/4}-N \right)}
 {\Gamma \left(N+\sqrt{N^2+1/4}-1/2 \right)} \right]^2 \\
 &\approx &
8\lambda N^2 \left(\frac{eh} {4\sqrt{2} N\lambda} \right)^{4N},
\end{eqnarray*}
where the last line is obtained under the limit $N \ll 1$. The
power law dependence on the field strength indicates that $\Delta
\varepsilon$ is vanishingly small for large $N$ unless the term
inside the bracket exceeds one, resulting in a critical field
strength at $(4\sqrt{2}/e) N\lambda \approx 0.7 h_c^{(c)}$.
Although this treatment explains the threshold behavior, it fails
to predict accurately the critical field strength. In particular,
it fails to predict the $N$-dependence of the ratio
$h_c^{(q)}/h_c^{(c)}$ [see Fig.~\ref{fig2}(b)] . This failure can
be easily understood since for $h$ close to the critical point, it
is no longer valid to regard the transverse field as a
perturbation.

\begin{figure}
\includegraphics*[width=7cm,height=7.5cm]{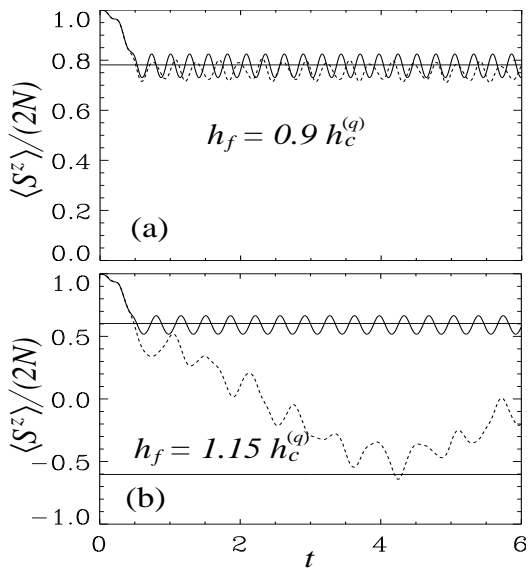}
\caption{Time evolution of the expectation value of $S^z$. Here
$N=10$. The solid lines represent the classical results and the
dashed lines the quantum mechanical results. The strength of the
transverse field, $h$, is ramped linearly from 0 to $h_f$ from
$t=0$ to $t=0.5$, then stays at $h_f$ afterwards. For (a),
$h_f=0.9 h_c$, no tunneling is present; for (b), $h_f =1.15h_c$,
macroscopic quantum coherence due to spin tunneling can be
observed in the quantum mechanical results. The horizontal lines
represent $S^z=\pm 2N \sqrt{1-[h_f/(3N \lambda)]^2}$, values of
$S^z$ in the classical ground state with $h=h_f$. The units for
time is $\hbar / \lambda$, which is on the order of a few seconds
for typical atomic parameters for pure magnetic dipolar
interaction. } \label{fig3}
\end{figure}

We now turn to the detection of tunneling. It is not practical to
directly measure the energy difference between the two lowest
energy states, since $\Delta \varepsilon$ is relatively small
compared to the total ground state energy. Instead, we use a
method which is perfectly suited for the situation at hand, where
thermal effects are negligible, but may not be appropriate for
more traditional solid-state systems. Our proposed detection
scheme starts with the system prepared in one of the degenerated
ground state, say, $|N,N \rangle$. The external field $h$ is then
slowly ramped up from zero to a final value, $h_f< h_c^{(c)}$.
Classically, the macroscopic spin would simply adiabatically
follow the instantaneous ground state closest to the initial
state. Quantum mechanically, though, this is only true as long as
$h_f < h_c^{(q)}$. But for $h_f > h_c^{(q)}$, the system is in the
quantum tunneling regime and does not reach an equilibrium.
Rather, it oscillates back and forth between the two classically
degenerate states, a phenomenon termed as {\em macroscopic quantum
coherence}. The observation of such oscillation will be a direct
signature of quantum tunneling, but it requires the effect of
dissipation to be small. Otherwise, spin relaxation will result,
which is normally the case for the experiment on solid magnetic
materials. The long coherence time associated with the atomic
condensate, however, makes it ideal for this purpose. In fact,
macroscopic center-of-mass oscillation has already been observed
in ultracold atoms trapped in optical lattices \cite{jessen}.
Fig.~\ref{fig3} depicts the numerical simulation of the
measurement scheme in both the classical and quantum treatments.
The quantum mechanical results are obtained by time evolving the
Schr\"{o}dinger equation in spin space with Hamiltonian
(\ref{hspin}), using the Crank-Nicolson method, while the
classical results are obtained by solving the dynamical equation
\begin{equation}
\frac{d{\bf S}_i}{dt} = - \gamma_B {\bf S}_i \times \frac{\delta
H}{\delta {\bf S}_i},
\end{equation}
where $H$ is the energy functional and has the same form as
Hamiltonian (\ref{hspin}) but with the spin operators treated as
$c$-numbers. Fig.~\ref{fig3}(a) is for $h_f < h_c^{(q)}$. There is
no qualitative difference between the quantum and classical
results, a steady state is reached in the end (small oscillations
around this steady state persists due to nonadiabiticity).
Fig.~\ref{fig3}(b) is for $h_f
> h_c^{(q)}$. The classical result is quite similar to that of
Fig.~\ref{fig3}(a). However, the quantum calculation clearly
shows the oscillations of the system between the two
macroscopically distinct states. These two states differ by a
minus sign in the expectation value of $S^z$, which can be easily
measured experimentally using, for example, Stern-Gerlach
technique \cite{jessen}.

In conclusion, we have shown that the quantum macroscopic
tunneling of spin is possible in a spinor condensate trapped in
double-well potential. Compared to the more conventional solid
state magnetic materials, our system possesses several decisive
advantages. First and perhaps foremost, this is an exceedingly
clean system characterized by a few simple parameters, without the
complication of domain separations, and with well-understood
microscopic physics; It is amenable to exquisite experimental
control; A single parameter --- the transverse field strength ---
is capable of switching on and off the tunneling. Typical
temperatures of the atomic condensate is in the nanoKelvin regime,
hence the thermal activation normally presented in the solid
materials can be safely neglected \cite{notespin}. Due to these
reasons, we believe that this system is indeed ideal for studying
quantum magnetism in general \cite{pu,spinwave}, and macrosopic
quantum tunneling in particular. Finally we remark that although
we have considered a double-well potential in this work, we expect
similar behavior for a condensate confined in a one-dimensional
periodic lattice potential.

We would like to thank Dr. Hui Hu for many valuable discussions
and correspondence, and Dr. Christian Tanguy for providing us the
final expression of $\Delta \varepsilon$ in the perturbation
treatment. This work is supported in part by the US Office of
Naval Research under Contract No. 14-91-J1205, by the National
Science Foundation under Grant No. PHY00-98129, by the US Army
Research Office, by NASA, and by the Joint Services Optics
Program.


\begin{references}

\bibitem{pu} H. Pu, W. Zhang and P. Meystre, Phys. Rev. Lett. {\bf
87}, 140405 (2001).

\bibitem{leggett}A. J. Leggett {\em et al.}, Rev. Mod. Phys. {\bf
59}, 1 (1987).

\bibitem{book} E. M. Chudnovsky, {\em Macroscopic Quantum
Tunneling of the Magnetic Moment} (Cambridge University Press,
1998).

\bibitem{qtm94} L. Gunther and B. Barbara, Eds., {\em Quantum Tunneling of
Magnetization - QTM'94} (Kluwer, Dordrecht, Netherlands, 1995).

\bibitem{exp}W. Wernsdorfer and R. Sessoli, Science {\bf 284}, 133
(1999); J. Brooke, T. F. Rosenbaum and G. Aeppli, Nature {\bf
413}, 610 (2001).

\bibitem{matrix} E. M. Chudnovsky and X. Martinez-Hidalgo, e-print
cond-mat/0201184.

\bibitem{milburn} G. J. Milburn {\em et al.}, Phys. Rev. A {\bf
55}, 4318 (1997).

\bibitem{smerzi}A. Smerzi, S. Fantoni, S. Giovanazzi and S. R.
Shenoy, Phys. Rev. Lett. {\bf 79}, 4950 (1997); S. Raghavan, A.
Smerzi, S. Fantoni and S. R. Shenoy, Phys. Rev. A {\bf 59}, 620
(1999).

\bibitem{jessen}D. L. Haycock {\em et al.}, Phys. Rev. Lett. {\bf
85}, 3365 (2000).

\bibitem{zhang}Y. Zhang and H. J. W. M\"{u}ller-Kirsten, Phys.
Rev. A {\bf 64}, 023608 (2001); L. D. Carr, K. W. Mahmud and W. P.
Reinhardt, Phys. Rev. A {\bf 64}, 033603 (2001); K. Kasamatsu, Y.
Yasui and M. Tsubota, Phys. Rev. A {\bf 64}, 053605 (2001).

\bibitem{law} C. K. Law, H. Pu and N. P. Bigelow, Phys. Rev. Lett.
{\bf 81}, 5257 (1998).

\bibitem{ho}T. -L. Ho, Phys. Rev. Lett. {\bf 81}, 742 (1998); T.
Ohmi and K. Machida, J. Phys. Soc. Jpn. {\bf 67}, 1882 (1998).

\bibitem{spinwave}W. Zhang, H. Pu, C. Search and P. Meystre, Phys.
Rev. Lett. {\bf 88}, 060401 (2002).

\bibitem{optical}S. Giovanazzi, D. O'Dell and G. Kurizki, Phys.
Rev. Lett. {\bf 88}, 130402 (2002).

\bibitem{note} For the very same reason, we have neglected
magnetic dipole-dipole interaction between atoms in the same well.

\bibitem{garg}A. Garg, Phys. Rev. B {\bf 60}, 6705 (1999).

\bibitem{qm}P. Roman, {\em Advanced Quantum Theory}
(Addison-Wesley, Reading, Massachusetts, 1995).

\bibitem{garanin}D. A. Garanin, J. Phys. A: Math. Gen. {\bf 24}, L61
(1991).\

\bibitem{notespin}In such a system, spin waves represent a kind of
excitation which can distort the ground state spin orientation
\cite{spinwave}. However, they will not be energetic enough to
induce tunneling.




\end{references}
\end{document}